\documentclass[twocolumn,11pt]{article}
\usepackage{amsmath,graphicx,url,amssymb,bbm}
\usepackage[left=1.7cm,right=1.7cm,bottom=2.4cm,top=2.1cm]{geometry}
\title{\textbf{\huge{Clotho: An Audio Captioning Dataset}}}
\author{Konstantinos Drossos, Samuel Lipping, and Tuomas Virtanen\vspace{9pt}\\
Audio Research Group, Tampere University, Tampere, Finland\\
\{firstname.lastname\}@tuni.fi
}
\date{}
\begin{document}
\twocolumn[
\maketitle
  \begin{@twocolumnfalse}
    \maketitle
    \begin{abstract}
    \normalsize
    Audio captioning is the novel task of general audio content description using free text. It is an intermodal translation task (not speech-to-text), where a system accepts as an input an audio signal and outputs the textual description (i.e. the caption) of that signal. In this paper we present Clotho, a dataset for audio captioning consisting of 4981 audio samples of 15 to 30 seconds duration and 24 905 captions of eight to 20 words length, and a baseline method to provide initial results. Clotho is built with focus on audio content and caption diversity, and the splits of the data are not hampering the training or evaluation of methods. All sounds are from the Freesound platform, and captions are crowdsourced using Amazon Mechanical Turk and annotators from English speaking countries. Unique words, named entities, and speech transcription are removed with post-processing. Clotho is freely available online (\url{https://zenodo.org/record/3490684}). 
    
    \vspace{6pt}
\end{abstract}
\vspace{18pt}
\end{@twocolumnfalse}
]
\section{Introduction}\label{sec:intro}
Captioning is the intermodal translation task of describing the human-perceived information in a medium, e.g. images (image captioning) or audio (audio captioning), using free text~\cite{lipping:2019:dcase,karpathy:2017:tpami,young:2014:tacl,drossos:2017:waspaa}. In particular, audio captioning was first introduced in~\cite{drossos:2017:waspaa}, it does not involve speech transcription, and is focusing on identifying the human-perceived information in an general audio signal and expressing it through text, using natural language. This information includes identification of sound events, acoustic scenes, spatiotemporal relationships of sources, foreground versus background discrimination, concepts, and physical properties of objects and environment. For example, given an audio signal, an audio captioning system would be able to generate captions like ``a door creaks as it slowly revolves back and forth''\footnote{Actual caption from the training split of Clotho dataset.\label{footnote:examples}}.

The dataset used for training an audio captioning method defines to a great extent what the method can learn~\cite{lipping:2019:dcase,yi:2014:coco}. Diversity in captions allows the method to learn and exploit the perceptual differences on the content (e.g. a thin plastic rattling could be perceived as a fire crackling)~\cite{lipping:2019:dcase}. Also, the evaluation of the method becomes more objective and general by having more captions per audio signal~\cite{yi:2014:coco}. 

Recently, two different datasets for audio captioning were presented, Audio Caption and AudioCaps~\cite{wu:2019:icassp,kim:2019:audiocaps}. Audio Caption is partially released, and contains 3710 domain-specific (hospital) video clips with their audio tracks, and annotations that were originally obtained in Mandarin Chinese and afterwards translated to English using machine translation~\cite{wu:2019:icassp}. The annotators had access and viewed the videos. The annotations contain description of the speech content (e.g. ``The patient inquired about the location of the doctor’s police station''). AudioCaps dataset has 46 000 audio samples from AudioSet~\cite{gemmeke:2017:icassp}, annotated with one caption each using the crowdsourcing platform Amazon Mechanical Turk (AMT) and automated quality and location control of the annotators~\cite{kim:2019:audiocaps}. Authors of AudioCaps did not use categories of sounds which they claimed that visuals were required for correct recognition, e.g. ``inside small room''. Annotators of AudioCaps were provided the word labels (by AudioSet) and viewed the accompanying videos of the audio samples. 

The perceptual ambiguity of sounds can be hampered by providing contextual information (e.g. word labels) to annotators, making them aware of the actual source and not letting them describe their own perceived information. Using visual stimuli (e.g. video) introduces a bias, since annotators may describe what they see and not what they hear. Also, a single caption per file impedes the learning and evaluation of diverse descriptions of information, and domain-specific data of previous audio captioning datasets have an observed significant impact on the performance of methods~\cite{wu:2019:icassp}. Finally, unique words (i.e. words appearing only once) affect the learning process, as they have an impact on the evaluation process (e.g. if a word is unique, will be either on training or on evaluation). An audio captioning dataset should at least provide some information on unique words contained in its captions. 

In this paper we present the freely available\footnote{\url{https://zenodo.org/record/3490684}\label{fn:clotho}}
audio captioning dataset Clotho\footnote{\url{https://en.wikipedia.org/wiki/Clotho}}, with 4981 audio samples and 24 905 captions. All audio samples are from Freesound platform~\cite{font:2013:freesound}, and are of duration from 15 to 30 seconds. Each audio sample has five captions of eight to 20 words length, collected by AMT and a specific protocol for crowdsourcing audio annotations, which ensures diversity and reduced grammatical errors~\cite{lipping:2019:dcase}. During annotation no other information but the audio signal was available to the annotators, e.g. video or word tags. The rest of the paper is organized as follows. Section~\ref{sec:data-collection} presents the creation of Clotho, i.e. gathering and processing of the audio samples and captions, and the splitting of the data to development, evaluation, and testing splits. Section~\ref{sec:baseline} presents the baseline method used, the process followed for its evaluation using Clotho, and the obtained results. Section~\ref{sec:conclusions} concludes the paper. 
%
%
\section{Creation of Clotho dataset}\label{sec:data-collection}
%
%
\subsection{Audio data collection and processing}
We collect the set of audio samples $\mathbb{X}_{\text{init}}=\{\mathbf{x}_{\text{init}}^{i}\}_{i=1}^{N_{\text{init}}}$, with $N_{\text{init}}=12000$ and their corresponding metadata (e.g. tags that indicate their content, and a short textual description), from the online platform Freesound~\cite{font:2013:freesound}. $\mathbf{x}_{\text{init}}$ was obtained by randomly sampling audio files from Freesound fulfilling the following criteria: lossless file type, audio quality at least 44.1 kHz and 16-bit, duration $10\text{ s}\leq d({\mathbf{x}_{\text{init}}^{i}})\leq300$ s (where $d(\mathbf{x})$ is the duration of $\mathbf{x}$), a textual description which first sentence does not have spelling errors according to US and UK English dictionaries (as an indication of the correctness of the metadata, e.g. tags), and not having tags that indicate music, sound effects, or speech. As tags indicating speech files we consider those like ``speech'', ``speak'', and ``woman''.  We normalize $\mathbf{x}^{i}_{\text{init}}$ to the range $[-1, 1]$, trim the silence (60 dB below the maximum amplitude) from the beginning and end, and resample to 44.1 kHz. Finally, we keep samples that are longer than 15 s as a result of the processing. This results in $\mathbb{X}'_{\text{init}}=\{\mathbf{x}_{\text{init}}^{j}\}_{j=1}^{N'_{\text{init}}},\,N'_{\text{init}}=9000$.

\begin{figure}[!t]
    \centering
    \includegraphics[width=.9\columnwidth,trim={0.1cm .25cm .4cm .4cm},clip]{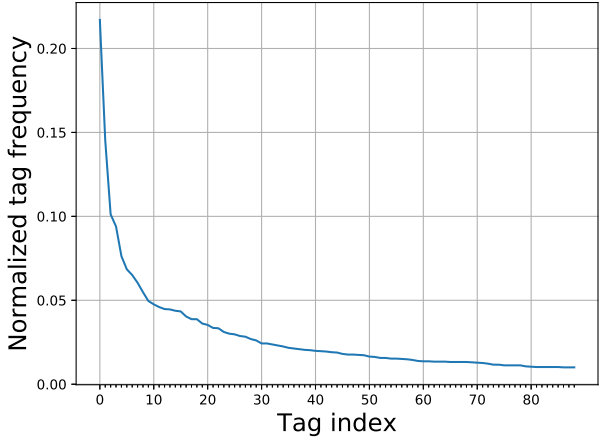}
    \caption{Distribution of tags in $\mathbb{T}_{\text{0.01}}$ for $\mathbb{X}_{\text{med}}$. Tags are sorted according to their frequency.}
    \label{fig:dist-tags}
\end{figure}

For enhancing the diversity of the audio content, we aim to create $\mathbb{X}_{\text{med}}\subset\mathbb{X}'_{\text{init}}$ based on the tags of $\mathbb{X}'_{\text{init}}$, targeting to the most uniform possible distribution of the tags of the audio samples in $\mathbb{X}_{\text{med}}$. We first create the bag of tags $\mathbb{T}$ by collecting all the tags of sounds in $\mathbb{X}'_{\text{init}}$. We omit tags that describe time or recording equipment and process (e.g. ``autumn'', ``field-recording''). Then, we calculate the normalized frequency of all tags in $\mathbb{T}$ and create $\mathbb{T}_{\text{0.01}}\subset\mathbb{T}$, with tags of a normalized frequency of at least 0.01. We randomly sample $10^6$ sets (with overlap) of $N_{\text{med}}=5000$ files from $\mathbb{X}'_{\text{init}}$, and keep the set that has the maximum entropy for $\mathbb{T}_{\text{0.01}}$. This process results in $\mathbb{X}_{\text{med}}=\{\mathbf{x}_{\text{init}}^{z}\}_{z=1}^{N_{\text{med}}}$, having the most uniform tag distribution and, hence, the most diverse content. The resulting distribution of the tags in $\mathbb{T}_{\text{0.01}}$ is illustrated in Figure~\ref{fig:dist-tags}. The 10 most common tags are: ambient, water, nature, birds, noise, rain, city, wind, metal, and people.

We target at audio samples $\mathbf{x}$ having a uniform distribution between 15 and 30 s. Thus, we further process $\mathbb{X}_{\text{med}}$, keeping the files with a maximum duration of 30 s and cutting a segment from the rest. We randomly select a set of values for the duration of the segments that will maximize the entropy of the duration of the files, discretizing the durations with a resolution of 0.05 s. In order to not pick segment without activity, we sample the files by taking a window with a selected duration that maximizes the energy of the sample. Finally, we apply a 512-point Hamming window to the beginning and the end of the samples, smoothing the effect of sampling. The above process results to $\mathbb{X}_{\text{sam}}=\{\mathbf{x}_{\text{sam}}^{z}\}_{z=1}^{N_{\text{med}}}$, where the distribution of durations is approximately uniform between 15 and 30 s.  
%
%
\subsection{Captions collection and processing}
We use AMT and a novel three-step based framework~\cite{lipping:2019:dcase} for crowdsourcing the annotation of $\mathbb{X}_{\text{sam}}$, acquiring the set of captions $\mathbb{C}_{\text{sam}}^{z}=\{c_{\text{sam}}^{z,u}\}_{u=1}^{N_{\text{cp}}}$ for each $\mathbf{x}_{\text{sam}}^{z}$, where $c_{\text{sam}}^{z,u}$ is an eight to 20 words long caption for $\mathbf{x}_{\text{sam}}^{z}$. In a nutshell, each audio sample $\mathbf{x}_{\text{sam}}^{z}$ gets annotated by $N_{\text{cp}}$ different annotators in the first step of the framework. The annotators have access only to $\mathbf{x}_{\text{sam}}^{z}$ and not any other information. In the second step, different annotators are instructed to correct any grammatical errors, typos, and/or rephrase the captions. This process results in $2\times N_{\text{cp}}$ captions per $\mathbf{x}_{\text{sam}}^{z}$. Finally, three (again different) annotators have access to $\mathbf{x}_{\text{sam}}^{z}$ and its $2\times N_{\text{cp}}$ captions, and score each caption in terms of the accuracy of the description and fluency of English, using a scale from 1 to 4 (the higher the better). The captions for each $\mathbf{x}_{\text{sam}}^{z}$ are sorted (first according to accuracy of description and then according to fluency), and two groups are formed: the top $N_{\text{cp}}$ and the bottom $N_{\text{cp}}$ captions. The top $N_{\text{cp}}$ captions are selected as $\mathbb{C}_{\text{sam}}^{z}$. We manually sanitize further $\mathbb{C}_{\text{sam}}^{z}$, e.g. by replacing ``it's'' with ``it is'' or ``its'', making consistent hyphenation and compound words (e.g. ``nonstop'', ``non-stop'', and ``non stop''), removing words or rephrasing captions pertaining to the content of speech (e.g. ``French'', ``foreign''), and removing/replacing named entities (e.g. ``Windex''). 

Finally, we observe that some captions include transcription of speech. To remove it, we employ extra annotators (not from AMT) which had access only at the captions. We instruct the annotators to remove the transcribed speech and rephrase the caption. If the result is less than eight words, we check the bottom $N_{\text{cp}}$ captions for that audio sample. If they include a caption that has been rated with at least 3 by all the annotators for both accuracy and fluency, and does not contain transcribed speech, we use that caption. Otherwise, we remove completely the audio sample. This process yields the final set of audio samples and captions, $\mathbb{X}=\{\mathbf{x}^{o}\}_{o=1}^{N}$ and $\mathbb{C}'=\{\mathbb{C}'^{o}\}_{o=1}^{N}$, respectively, with $\mathbb{C}'^{o}=\{c'^{o,u}\}_{u=1}^{N_{\text{cp}}}$ and $N=4981$.

An audio sample should belong to only one split of data (e.g., training, development, testing). This means that if a word appears only at the captions of one $\mathbf{x}^{o}$, then this word will be appearing only at one of the splits. Having a word appearing only in training split leads to sub-optimal learning procedure, because resources are spend to words unused in validation and testing. If a word is not appearing in the training split, then the evaluation procedure suffers by having to evaluate on words not known during training. For that reason, for each $\mathbf{x}^{o}$ we construct the set of words $\mathbb{S}_{a}^{o}$ from $\mathbb{C}'^{o}$. Then, we merge all $\mathbb{S}_{a}^{o}$ to the bag $\mathbb{S}_{T}$ and we identify all words that appear only once (i.e. having a frequency of one) in $\mathbb{S}_{T}$. We employ an extra annotator (not from AMT) which has access only to the captions of $\mathbf{x}^{o}$, and has the instructions to change the all words in $\mathbb{S}_{T}$ with frequency of one, with other synonym words in $\mathbb{S}_{T}$ and (if necessary) rephrase the caption. The result is the set of captions $\mathbb{C}=\{\mathbb{C}^{o}\}_{o=1}^{N}$, with words in $\mathbb{S}_{T}$ having a frequency of at least two. Each word will appear in the development set and at least in one of the evaluation or testing splits. This process yields the data of the Clotho dataset, $\mathbb{D}=\{\left<\mathbf{x}^{o}, \mathbb{C}^{o}\right>\}_{o=1}^{N}$.
%
%
\subsection{Data splitting}\label{sec:data-split}
We split $\mathbb{D}$ in three non-overlapping splits of 60\%-20\%-20\%, termed as development, evaluation, and testing, respectively. Every word in the captions of $\mathbb{D}$ appears at the development split and at least in one of the other two splits. 

\begin{figure}
    \centering
    \includegraphics[width=.9\columnwidth,trim={0.1cm .25cm .4cm .4cm},clip]{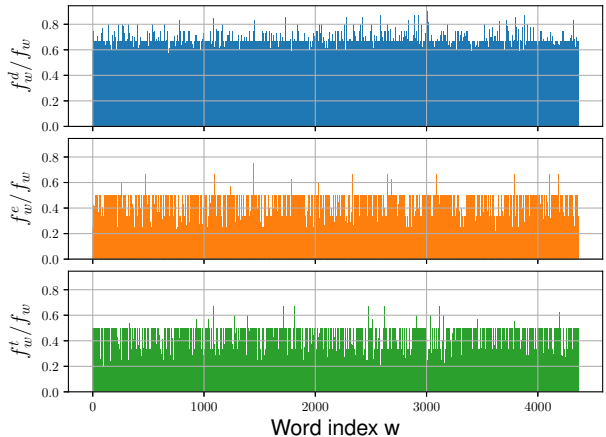}
    \caption{Percentage of words ($f^{d}_{w}/f_{w}$, $f^{e}_{w}/f_{w}$, and $f^{t}_{w}/f_{w}$) in the three different splits}
    \label{fig:perc-splits}
\end{figure}

For each $\mathbf{x}^{o}$ we construct the set of unique words $\mathbb{S}^{o}$ from its captions $\mathbb{C}^{o}$, using all letters in small-case and excluding punctuation. We merge all $\mathbb{S}^{o}$ to the bag $\mathbb{S}_{\text{bag}}$ and calculate the frequency $f_{w}$ of each word $w$. We use multi-label stratification\footnote{\url{https://github.com/trent-b/iterative-stratification}}~\cite{sechidis:2011:mlkdd}, having as labels for each $\mathbf{x}^{o}$ the corresponding words $\mathbb{S}^{o}$, and split $\mathbb{D}$ 2000 times in sets of splits of 60\%-40\%, where 60\% corresponds to the development split. We reject the sets of splits that have at least one word appearing only in one of the splits. Ideally, the words with $f_{w}=2$ should appear only once in the development split. The other appearance of word should be in the evaluation or testing splits. This will prevent having unused words in the training (i.e. words appearing only in the development split) or unknown words in the evaluation/testing process (i.e. words not appearing in the development split). The words with $f_{w}\geq3$ should appear $f^{\text{Dev}}_{w}=\lfloor0.6f_{w}\rfloor$ times in the development split, where 0.6 is the percentage of data in the development split and $\lfloor\ldots\rfloor$ is the floor function. We calculate the frequency of words in the development split, $f^{\text{d}}_{w}$, and observe that it is impossible to satisfy the $f^{\text{d}}_{w}=f^{\text{Dev}}_{w}$ for the words with $f_{w}\geq3$. Therefore, we adopted a tolerance $\delta_{w}$ (i.e. a deviation) to the $f^{\text{Dev}}_{w}$, used as $f^{\text{Dev}}_{w}\pm\delta_{w}$:
\begin{equation}\label{eq:tolerance}
    \delta_{w}=\begin{cases}
    1&\text{if }f_{w}\in[3,6]\text{,}\\
    2&\text{if }f_{w}\in[7,16]\text{,}\\
    4&\text{if }f_{w}\in[17,20]\text{,}\\
    \lfloor0.2f_{w}\rfloor&\text{otherwise.}
    \end{cases}
\end{equation}
\noindent
The tolerance means, for example, that we can tolerate a word appearing a total of 3 times in the whole Clotho dataset $\mathbb{D}$, to appear 2 times in the development split (appearing 0 times in development split results in the rejection of the split set). This will result to this word appearing in either evaluation or testing split, but still this word will not appear only in one split. To pick the best set of splits, we count the amount of words that have a frequency $f^{\text{d}}_{w}\notin[f^{\text{Dev}}_{w}-\delta_{w},f^{\text{Dev}}_{w}+\delta_{w}]$. We score, in an ascending fashion, the sets of splits according to that amount of words and we pick the top 50 ones. 
For each of the 50 sets of splits, we further separate the 40\% split to 20\% and 20\%, 1000 times. That is, we end up with 50 000 sets of splits of 60\%, 20\%, 20\%, corresponding to development, evaluation, and testing splits, respectively. We want to score each of these sets of splits, in order to select the split with the smallest amount of words that deviate from the ideal split for each of these 50 000 sets of splits. We calculate the frequency of appearance of each word in the development, evaluation, and testing splits, $f^{\text{d}}_{w}$, $f^{\text{e}}_{w}$, and $f^{\text{t}}_{w}$, respectively. Then, we create the sets of words $\Psi_{d}$, $\Psi_{e}$, and $\Psi_{t}$, having the words with $f^{\text{d}}_{w} \notin[f^{\text{Dev}}_{w}- \delta_{w},f^{\text{Dev}}_{w}+\delta_{w}]$, $f^{\text{e}}_{w} \notin[f^{\text{Ev}}_{w}- \delta_{w},f^{\text{Ev}}_{w}+\delta_{w}]$, and $f^{\text{t}}_{w} \notin[f^{\text{Ev}}_{w}- \delta_{w},f^{\text{Ev}}_{w}+\delta_{w}]$, respectively, where $f^{\text{Ev}}_{w} = f_{w} - f^{\text{Dev}}_{w}$. Finally, we calculate the sum of the weighted distance of frequencies of words from the $f^{\text{Dev}}_{w}\pm\delta_{w}$ or $f^{\text{Ev}}_{w}\pm\delta_{w}$ range (for words being in the development split or not, respectively), $\Gamma$, as
\begin{align}
    \Gamma = &\sum_{w\in\Psi_{d}}(\alpha_{d}|f^{\text{Dev}}_{w} - f^{d}_{w}| - \delta_{w}) + \sum_{w\in\Psi_{e}}(\alpha_{e}|f^{\text{Ev}}_{w} - f^{e}_{w}| - \delta_{w})\nonumber\\
    +&\sum_{w\in\Psi_{t}}(\alpha_{e}|f^{\text{Ev}}_{w} - f^{t}_{w}| - \delta_{w})
\end{align}
\noindent
where $\alpha_{d}=1/f^{\text{Dev}}_{w}$ and $\alpha_{e}=1/0.5f^{\text{Ev}}_{w}$. We sort all 50 000 sets of splits according to $\Gamma$ and in ascending fashion, and we pick the top one. This set of splits is the final split for the Clotho dataset, containing 2893 audio samples and 14465 captions in development split, 1045 audio samples and 5225 captions in evaluation split, and 1043 audio samples and 5215 captions in the testing split. The development and evaluation splits are freely available online\textsuperscript{2}. The testing split is withheld for potential usage in scientific challenges. A fully detailed description of the Clotho dataset can be found online\footnote{\url{url-to-be-announced}}. In Figure~\ref{fig:perc-splits} is a histogram of the percentage of words ($f^{d}_{w}/f_{w}$, $f^{e}_{w}/f_{w}$, and $f^{t}_{w}/f_{w}$) in the three different splits.
%
%
\section{Baseline method and evaluation}\label{sec:baseline}
In order to provide an example of how to employ Clotho and some initial (baseline) results, we use a previously utilized method for audio captioning~\cite{drossos:2017:waspaa} which is based on an encoder-decoder scheme with attention. The method accepts as an input a length-$T$ sequence of 64 log mel-band energies $\mathbf{X}\in\mathbb{R}^{T\times64}$, which is used as an input to a DNN which outputs a probability distribution of words. The generated caption is constructed from the output of the DNN, as in~\cite{drossos:2017:waspaa}. We optimize the parameters of the method using the development split of Clotho and we evaluate it using the evaluation and the testing splits, separately. 

We first extract 64 log mel-band energies, using a Hamming window of 46 ms, with 50\% overlap. We tokenize the captions of the development split, using a one-hot encoding of the words. Since all the words in in the development split appear in the other two splits as well, there are no unknown tokens/words. We also employ the start- and end-of-sequence tokens ($\left<\text{SOS}\right>$ and $\left<\text{EOS}\right>$ respectively), in order to signify the start and end of a caption. 

The encoder is a series of bi-directional gated recurrent units (bi-GRUs)~\cite{cho:2014:emnlp}, similarly to~\cite{drossos:2017:waspaa}. The output dimensionality for the GRU layers (forward and backward GRUs have same dimensionality) is $\{256, 256, 256\}$. The output of the encoder is processed by an attention mechanism and its output is given as an input to the decoder. The attention mechanism is a feed-forward neural network (FNN) and the decoder a GRU. Then, the output of the decoder is given as an input to another FNN with a softmax non-linearity, which acts as a classifier and outputs the probability distribution of words for the $i$-th time-step. To optimize the parameters of the employed method, we use five times each audio sample, using its five different captions as targeted outputs each time. We optimize jointly the parameters of the encoder, attention mechanism, decoder, and the classifier, using 150 epochs, the cross entropy loss, and Adam optimizer~\cite{kingma:2015:adam} with proposed hyper-parameters. Also, in each batch we pad the captions of the batch to the longest in the same batch, using the end-of-sequence token, and the input audio features to the longest ones, by prepending zeros.

We assess the performance of the method on evaluation and testing splits, using the machine translation metrics BLEU\textsubscript{$n$} (with $n=1,\ldots,4$), METEOR, CIDEr, and ROUGE\textsubscript{L} for comparing the output of the method and the reference captions for the input audio sample. In a nutshell, BLEU\textsubscript{$n$} measures a modified precision of $n$-grams (e.g. BLEU\textsubscript{2} for 2-grams), METEOR measures a harmonic mean-based score of the precision and recall for unigrams, CIDEr measures a weighted cosine similarity of $n$-grams, and ROUGE\textsubscript{L} is a longest common subsequence-based score.

In Table~\ref{tab:metric-table} are the scores of the employed metrics for the evaluation and testing splits. 
\begin{table}[t]
\centering
\caption{Translation metrics for the evaluation and testing splits. B\textsubscript{$n$}, C, M, and R correspond to BLEU\textsubscript{$n$}, CIDEr, METEOR, and ROUGE, respectively.}
\label{tab:metric-table}
\bigskip
\resizebox{\columnwidth}{!}{
\begin{tabular}{lccccccc}
\textbf{Metric} & \textbf{B\textsubscript{1}} & \textbf{B\textsubscript{2}} & \textbf{B\textsubscript{3}} & \textbf{B\textsubscript{4}} & \textbf{C} & \textbf{M} & \textbf{R}\\ \hline
Evaluation & 0.42 & 0.14 & 0.06 & 0.02 & 0.10 & 0.09 & 0.27 \\ 
Testing    & 0.42  & 0.14  & 0.05   & 0.02 & 0.09 & 0.08 & 0.26 
\end{tabular}\bigskip}
\end{table}
As can be seen from Table~\ref{tab:metric-table} and BLEU\textsubscript{1}, the method has started identifying the content of the audio samples by outputting words that exist in the reference captions. For example, the method outputs ``water is running into a container into a'', while the closest reference caption is ``water pouring into a container with water in it already'', or ``birds are of chirping the chirping and various chirping'' while the closest reference is ``several different kinds of birds are chirping and singing''. The scores of the rest metrics reveal that the structure of the sentence and order of the words are not correct. These are issues that can be tackled by adopting either a pre-calculated or jointly learnt language model. In any case, the results show that the Clotho dataset can effectively be used for research on audio captioning, posing useful data in tackling the challenging task of audio content description. 
%
%
\section{Conclusions}
\label{sec:conclusions}
In this work we present a novel dataset for audio captioning, named Clotho, that contains 4981 audio samples and five captions for each file (totaling to 24 905 captions). During the creating of Clotho care has been taken in order to promote diversity of captions, eliminate words that appear only once and named entities, and provide data splits that do not hamper the training or evaluation process. Also, there is an example of the usage of Clotho, using a method proposed at the original work of audio captioning. The baseline results indicate that the baseline method started learning the content of the input audio, but more tuning is needed in order to express the content properly. Future work includes the employment of Clotho and development of novel methods for audio captioning.
%
%
\section*{Acknowledgement}
The research leading to these results has received funding from the European Research Council under the European Union’s H2020 Framework Programme through ERC Grant Agreement 637422 EVERYSOUND. Part of the computations leading to these results were performed on a TITAN-X GPU donated by NVIDIA to K. Drossos. The authors also wish to acknowledge CSC-IT Center for Science, Finland, for computational resources.
%
%
\bibliographystyle{IEEEbib}
\bibliography{refs}

\end{document}